\documentclass{IEEEtran4PSCC}

\usepackage{cite}

\ifCLASSINFOpdf
   \usepackage[pdftex]{graphicx}
\else
   \usepackage[dvips]{graphicx}
\fi
\usepackage[cmex10]{amsmath}
\usepackage{url}

\usepackage{color}

\DeclareMathOperator*{\maxmin}{max\,min}

\usepackage{lscape}

\usepackage{framed}

\newenvironment{mydescription}[1]
  {\begin{list}{}%
   {\renewcommand\makelabel[1]{##1 \hfill}%
   \settowidth\labelwidth{\makelabel{#1}}%
   \setlength\leftmargin{\labelwidth}
   \addtolength\leftmargin{\labelsep}}}
  {\end{list}}

\makeatletter
\let\old@ps@headings\ps@headings
\let\old@ps@IEEEtitlepagestyle\ps@IEEEtitlepagestyle
\def\psccfooter#1{%
    \def\ps@headings{%
        \old@ps@headings%
        \def\@oddfoot{\strut\hfill#1\hfill\strut}%
        \def\@evenfoot{\strut\hfill#1\hfill\strut}%
    }%
    \def\ps@IEEEtitlepagestyle{%
        \old@ps@IEEEtitlepagestyle%
        \def\@oddfoot{\strut\hfill#1\hfill\strut}%
        \def\@evenfoot{\strut\hfill#1\hfill\strut}%
    }%
    \ps@headings%
}
\makeatother

\psccfooter{%
}

\begin{document}
%
\title{Cyber-physical risk modeling with imperfect cyber-attackers \vspace{-0.2in}}

\author{
\IEEEauthorblockN{Efthymios Karangelos and Louis Wehenkel,}
\IEEEauthorblockA{Department of EE \& CS - Montefiore Institute, \\
University of Li\`{e}ge, Belgium.\\
\{e.karangelos;l.wehenkel\}@uliege.be}

}


\maketitle
 
\begin{abstract} 
We model the risk posed by a malicious cyber-attacker seeking to induce grid insecurity by means of a \emph{load redistribution} attack, while explicitly acknowledging that such an actor would plausibly base its decision strategy on imperfect information. More specifically, 
we introduce a novel formulation for the cyber-attacker's decision-making problem and analyze the distribution of decisions taken with randomly inaccurate data on the grid branch admittances or capacities, and the distribution of their respective impact. Our findings indicate that inaccurate admittance values most often lead to suboptimal cyber-attacks that still compromise the grid security, while inaccurate capacity values result in notably less effective attacks. 
We also find  common attacked cyber-assets and common affected physical-assets between all (random) imperfect cyber-attacks, which could be exploited in a preventive and/or corrective sense for effective cyber-physical risk management.
\end{abstract}


\thanksto{\noindent This work has been prepared with the support of the Belgian Energy Transition Fund,
project CYPRESS (\url{https://cypress-project.be/}).}

\section{Introduction}

The digitalization of  electric power system control \& communications is reshaping the scope for security management. In addition to \emph{physical} threats (\textit{e.g.}, failures of the physical infrastructure, forecasting errors, \textit{etc.}), securing the system against \emph{cyber} threats has also become essential \cite{kirschen2008}. Such  {threats} notably includes the adversary actions of malicious external agents, seeking to exploit weaknesses in the security of the system cyberspace so as to disrupt the physical supply of electricity. Going from physical to cyber-physical security management therefore requires modeling not only the interdependencies between cyber and physical infastructure {s}, but also the interaction between \emph{cyber-attackers} and  \emph{grid-operator {s}}.

\subsection{Related literature}
The framework of multi-level optimization is most commonly used in the state of the art to model the cyber-attacker \textit{vs} grid-operator interaction \cite{Zhang2021,he2016}.  The foundations of this literature lie in \cite{liu2011}, showing that an attacker can successfully introduce false data without being detected in the case of linear state estimation. Accordingly, bilevel formulations including a malicious attacker as the upper agent and a grid-operator solving the linear \emph{DC Optimal Power Flow} (DC-OPF) problem have been proposed to model alternative cyber-physical attack scenarios. Yuan \textit{et al.} introduced  {in  \cite{yuan2011}} the concept of \emph{load redistribution} while modeling an attacker seeking to  falsify load measurements so as to provoke  out-of-merit generation dispatching by the grid-operator. The same concept was exploited in \cite{liang2015} by an attacker seeking to maximize the loading of a transmission line.  Zhang and Sankar developed  {in  \cite{zhang2016}} an elaborate bilevel formulation for an attacker using load redistribution in order to hide a physical change in the grid topology, while a model of an attacker seeking to maximize the number of overloads induced through load redistribution has been presented in \cite{tian2019}.
  
We underline the  shared assumption of a cyber-attacker relying on \emph{perfect} information (\textit{i.e.}, the correct model of the power system and of its operator) to determine its attack strategy. However, Rahman and Mohsenian-Rad \cite{rahman2012}  stress that realistic \emph{imperfect} cyber-attackers would have to rely on an inaccurate grid model, as they cannot be plausibly assumed to observe in real-time the status of every circuit breaker, tap-changer  \textit{etc.}.  These authors modeled an attacker with incorrect line admittance data and found that the probability of detecting a load redistribution attack designed with  such imperfect information remains rather low.  Reference \cite{zhang2018} provides further evidence for challenging the perfect information assumption by means of a sensitivity analysis with respect to the attacker's knowledge of the occurrence of a single line outage, showing that the incorrect grid topology  undermines the evaluation of the cyber-physical attack.  From a game theory perspective, Sanjab and Saad studied  {in \cite{anibal}} the interaction between a defender taking preventive actions and potential realistic cyber-attackers with limited system knowledge and found that the Nash-equilibrium strategy against the assumed perfect, fully rational attacker is not the best defense of the grid.

\subsection{Paper scope \& contributions}

In this paper we focus on cyber-physical risk modeling while explicitly acknowledging that a realistic cyber-attacker would plausibly base her strategy on imperfect information.  We investigate the effect of such imperfect information in terms of: i) the impact of a cyber-physical attack on the electricity transmission grid and ii) the various attack vectors that may  be launched by a malicious cyber-attacker. The former relates primarily to cyber-physical risk assessment applications, while the latter allows to draw conclusions for cyber-physical risk management. Our analysis is based on the well-studied scenario of load redistribution attacks wherein a malicious attacker seeks to deceive the grid-operator with the final purpose of rendering the grid insecure. On top of imperfect information on the network branch admittances (as in \cite{rahman2012}), we posit that a realistic cyber-attacker may also rely on inaccurate branch capacity data. Indeed, the ability of transmission branches to securely sustain loading relies on ambient conditions and so does in practice the tolerance of grid-operators for higher loading levels  {when these ambient conditions are favourable}.  

To perform our investigation we use the \emph{Monte Carlo}  framework while modeling sequentially the decisions of a (random) imperfect cyber-attacker, the corresponding reaction of the grid-operator and finally the resulting state of the grid. The core component of our framework is a bilevel   {``$\maxmin$''} optimization model of a cyber-attacker, anticipating the reaction of the grid-operator to a load redistribution attack.  In addition to the standard constraints included in the  {state of the art} of such models, we formalize here novel constraint expressions to reflect a malicious cyber-attacker with the intention of inducing  a challenging insecure state,  potentially triggering a cascading failure event.

Our analysis showcases that (even minor) informational imperfections imply a broad spectrum of potential cyber-attacks and of respective physical impacts on the electricity system.  Moreover, the spectrum of potential cyber-attacks clearly features groups of common  attacked assets in the cyber sub-system and common affected assets in the physical sub-system. The implication is that protecting the cyber sub-system to avoid/detect intrusion of such common  attacked assets and/or the physical sub-system to withstand the possible failure of such common affected assets  {could} be effective cyber-physical risk management strategies.

\subsection{Paper organization}

Section \ref{blv} introduces the model of a malicious cyber-attacker seeking to maximize grid insecurity on the basis of the data  perceives to be true. Section \ref{method} presents the proposed methodology and metrics for acknowledging the informational imperfections of such an actor in the context of risk assessment and risk management. Section \ref{casestudy} discusses the application of such methodology on the single area version IEEE-RTS96 benchmark \cite{test_system} while conclusions are drawn in Section \ref{theend}.

\section{Cyber-attacker decision-making problem}
\label{blv}

We model a malicious cyber-attacker seeking to maximize the grid physical insecurity through a load redistribution attack.
More specifically, we consider an attacker falsifying bus load measurements  so as to mislead the grid-operator into perceiving the grid as insecure and implementing unnecessarily generation redispatch actions. The cyber-attacker's objective is to maximize the total magnitude of branch overloads caused by the injection of false measurements and the resulting generation redispatching of the mislead grid-operator. Invoking the DC-power flow approximation, we cast the cyber-attacker's decision making problem as the following bi-level \emph{Mixed Integer Linear Programming} (MILP) problem:
\begin{flalign}
\max&\sum_{\ell \in \mathcal{L}}r_{\ell}&\label{up.obj}  \\
&\sum_{\ell \in \mathcal{L}} \left(u_{\ell}^{+}+  u_{\ell}^{-}\right) \ge U & \label{min_order}\\
&\sum_{n \in \mathcal{N}} a_{n}\le A &\label{fd_budget}\\
&\sum_{n \in \mathcal{N}} e_{n}=0 &\label{fd_balance}\\
\intertext{\textit {for all nodes} $n \in \mathcal{N}$:}
&-a_n\cdot \epsilon \cdot d_{n} \le e_n \le a_n\cdot \epsilon \cdot d_{n}& \label{fd_perbus}\\
&a_n \in \{0,1\}& \label{fd_bin}\\
&\sum_{g \in \mathcal{G}} \gamma_{g,n}\left(p_{g0}+p_g^{\star}\right)-\sum_{\ell \in \mathcal{L}}\lambda_{\ell,n}\cdot f_{\ell}^{ca}=d_{n}& \label{true_pb}\\
\intertext{\textit {for all branches} $\ell \in \mathcal{L}$:}
&f_{\ell}^{ca}=\left(1/X_{\ell}\right)\cdot \sum_{n \in \mathcal{N}}\lambda_{\ell,n}\cdot \theta_{n}^{ca}& \label{true_flows}\\
&u_{\ell}^{+}+  u_{\ell}^{-}+u_{\ell}^{0}= 1& \label{flowflags}\\
& f_{\ell}^{ca}-\rho_{\ell}\cdot \overline{f}_{\ell}\le u_{\ell}^{+}\cdot M & \label{flagbins1}\\
&f_{\ell}^{ca}-\rho_{\ell}\cdot \overline{f}_{\ell}\ge (u_{\ell}^{+}-1)\cdot M& \label{flagbins2}\\
& -f_{\ell}^{ca}-\rho_{\ell}\cdot \overline{f}_{\ell}\le u_{\ell}^{-}\cdot M & \label{flagbins3}\\
&f_{\ell}^{ca}+\rho_{\ell}\cdot \overline{f}_{\ell}\ge (1-u_{\ell}^{-})\cdot M& \label{flagbins4}\\
&r_{\ell}\le (1-u_{\ell}^{0})\cdot M & \label{flagbins5}\\
&(u_{\ell}^{+}-1)\cdot M + (f_{\ell}^{ca}-\overline{f}_{\ell})\le r_{\ell} &\label{flagbins6}\\
&r_{\ell}\le (1-u_{\ell}^{+})\cdot M + (f_{\ell}^{ca}-\overline{f}_{\ell})&\label{flagbins7}\\
&(u_{\ell}^{-}-1)\cdot M - (f_{\ell}^{ca}+\overline{f}_{\ell})\le r_{\ell} &\label{flagbins8}\\
&r_{\ell}\le (1-u_{\ell}^{-})\cdot M - (f_{\ell}^{ca}+\overline{f}_{\ell})&\label{flagbins9}\\
&u_{\ell}^{+}, u_{\ell}^{-},u_{\ell}^{0}\in \{0,1\}& \label{flagbins10}\\
\intertext{ {\textit {subject to the model of the mislead grid-operator}:}} 
&\min\sum_{g \in \mathcal{G}}c_g\cdot \pi_g&\label{low.obj}\\
\intertext{\hspace{10mm} \small{\textit {for all generators} $g \in \mathcal{G}$:}}
&\hspace{8mm}  \pi_{g}\ge 0&\label{pcostnz}\\
&\hspace{8mm}  \pi_{g}\ge p_{g}^{\star}&\label{pcost}\\
&\hspace{8mm}(\underline{p}_{g}-p_{g0})\le p_{g}^{\star}\le (\overline{p}_g-p_{g0})&\label{pgen}\\
\intertext{\hspace{10mm} \small{\textit {for all nodes} $n \in \mathcal{N}$:}}
&\hspace{8mm} \sum_{g \in \mathcal{G}} \gamma_{g,n}\left(p_{g0}+p_g^{\star}\right)\hspace{-0.5mm}-\hspace{-0.5mm}\sum_{\ell \in \mathcal{L}}\lambda_{\ell,n}f_{\ell}^{go}\hspace{-0.5mm}=\hspace{-0.5mm}d_{n}+e_n& \label{lowpb}\\
\intertext{\hspace{10mm} \small{\textit {for all branches} $\ell \in \mathcal{L}$:}}
&\hspace{8mm} f_{\ell}^{go}=\left(1/X_{\ell}\right)\cdot \sum_{n \in \mathcal{N}}\lambda_{\ell,n}\cdot \theta_{n}^{go}& \label{lowflows}\\
&\hspace{10mm} -\overline{f}_{\ell}\le f_{\ell}^{go} \le \overline{f}_{\ell}.&\label{prat}
\end{flalign}

\begin{mydescription}{$sev$}
	\item[$\mathcal{G}$]{set of generating units;}
	\item[$\mathcal{L}$]{set of transmission branches;}
	\item[$\mathcal{N}$]{set of nodes;}
	\item[$r_{\ell}$]{upper-level continuous variable, measuring the magnitude of the branch overloads induced by the attack;} 
	\item[$u_{\ell}^{\cdot}$]{upper-level binary variable, indicating the overload status of branch $\ell$, with superscripts $(+/-)$ for an overloaded branch in the positive/negative flow direction or $0$ for no overload;}
	\item[$U$]{parameter, modeling the minimum number of overloaded branches targeted by the attacker;}
	\item[$a_{n}$]{upper-level binary variable, indicating the injection of false data at the load measurement of node $n$;}		
	\item[$A$]{parameter, modeling the attacker's available budget for attacking the grid load meters;}
	\item[$e_{n}$]{upper-level continuous variable, modeling the false active power demand measurement data injected by the attacker at node $n$;}	
	\item[$\epsilon$]{parameter, modeling the maximum relative amount of false load measurement data that can be injected by the attacker;}
	\item[$d_n$]{parameter, modeling the active power demand at node $n$;}
	\item[$\gamma_{g,n}$]{parameter, modeling the connectivity of generator $g$ with node $n$;}
	\item[${p}_{g0}$]{parameter, modeling the dispatch of generator $g$;}
	\item[$p_g$]{lower-level continuous variable, modeling the  active power redispatch of generator $g$ by the grid-operator;}
	\item[$\lambda_{\ell,n}$]{parameter, modeling the connectivity of branch $\ell$ with node $n$ and the assumed flow direction;}
	\item[$f_{\ell}^{ca}$]{upper-level continuous variable, modeling the cyber-attacker's perceived active power flow  value through branch $\ell$;}
	\item[$X_{\ell}$]{parameter, modeling the reactance of branch $\ell$;}
	\item[$\theta_n^{ca}$]{upper-level continuous variable, modeling the cyber-attacker's perceived voltage angle value at node $n$;}
	\item[$\rho_{\ell}$]{parameter, modeling the minimum threshold of overloaded flow per branch targeted by the attacker;}
	\item[$\overline{f}_{\ell}$]{parameter, modeling the capacity of branch $\ell$;}
	\item[$M$]{a large constant parameter;}
	\item[${c}_g$]{parameter, modeling the non-negative upward redispatch marginal cost of generator $g$;}
	\item[$\pi_g$]{lower-level continuous variable, modeling the upward redispatch of generator $g$;} 
	\item[$\underline{p}_g$]{parameter, modeling the minimum stable output of generator $g$;}
	\item[$\overline{p}_g$]{parameter, modeling the capacity of generator $g$;}
	\item[$f_{\ell}^{go}$]{lower-level continuous variable, modeling the grid-operator's perceived active power flow value through branch $\ell$;}
	\item[$\theta_n^{go}$]{upper-level continuous variable, modeling the grid-operator's voltage angle value at node $n$.}
\end{mydescription}

Objective function \eqref{up.obj} seeks to maximize the total magnitude of the branch overloads induced by the cyber-attack. We introduce inequality constraint \eqref{min_order} to model that a malicious cyber-attacker may strategically prefer to overload at least a minimum number of branches ($U\ge 2$) in order to create an overwhelming grid insecurity instance outside the \emph{``comfort zone''} of N-1 security.

Expression \eqref{fd_budget} imposes a limit on the maximum number of load meters that can be manipulated by the attacker, while \eqref{fd_balance} enforces that the false load measurement data injection is balanced across the grid and \eqref{fd_perbus} sets the maximum relative amount of false data that can be injected by the attacker at any node\footnote{As discussed in \cite{che2018false} constraints (\ref{fd_balance},\ref{fd_perbus}) are the standard \emph{proxy} constraints for the undetectability of a load redistribution attack in the DC model.}. Equalities (\ref{true_pb},\ref{true_flows}) model the power flow of the grid as perceived by the cyber-attacker only. Notice that the power balance constraint \eqref{true_pb} includes the optimal values of the generation redispatch variables $(p_{g}^{\star})$ as decided in the grid-operator's lower-level problem (\ref{low.obj} -- \ref{prat}). 

The group of inequalities (\ref{flagbins1} -- \ref{flagbins4}) is used to flag overloaded branches either in the positive or in the negative flow direction, while  (\ref{flagbins5} -- \ref{flagbins9}) are used to measure the magnitude of the branch overloads caused by the cyber attack. Here we originally  introduce a parameter $(\rho_{\ell}\ge 1)$  to model that a malicious cyber-attacker may strategically prefer to cause an overloaded flow larger than a threshold on every overloaded branch in order to create an overwhelming grid insecurity instance. Indeed, by way of  (\ref{flagbins1} -- \ref{flagbins9}), only overloads above such threshold contribute in the right-hand-side of constraint \eqref{min_order}  and  objective function \eqref{up.obj}.

The lower-level problem (\ref{low.obj} -- \ref{prat}) is a standard DC-OPF problem modeling the reaction of the grid-operator to the injection of the false data by the attacker, seeking to minimize the cost of upward generation redispatching so as to maintain all perceived (\textit{i.e.}, false) branch flow values within the respective capacity ratings. We note here the addition of the false data injection variable $(e_n)$ in the right-hand-side of the power balance constraint \eqref{lowpb}, as well as the use of supersrcipt $(^{go})$ to denote the (false) branch flow and voltage angle values perceived by the grid-operator.
 
\color{black}

\section{Models \& metrics for cyber-attacks with imperfect information}
\label{method}
We follow the Monte Carlo approach while sampling random error terms for the grid parameters to reflect that a cyber-attacker with imperfect information would base  {her} decisions on randomly inaccurate grid parameter values. More specifically, we assume that the branch admittances or transmission capacities may be imperfectly known by the attacker and form a simulation sample by drawing a unique, uniformly distributed, error term per branch.  For each simulation sample, we model the sequence of a cyber-attack as detailed in \ref{attack_model}. To analyze the resulting distribution of cyber-attacks in the context of cyber-physical risk assessment and risk management we introduce novel metrics in sections \ref{assess_metrics} and \ref{mgmt_metrics} respectively.


\begin{figure}[h]
\centering
\includegraphics[width=0.45\textwidth]{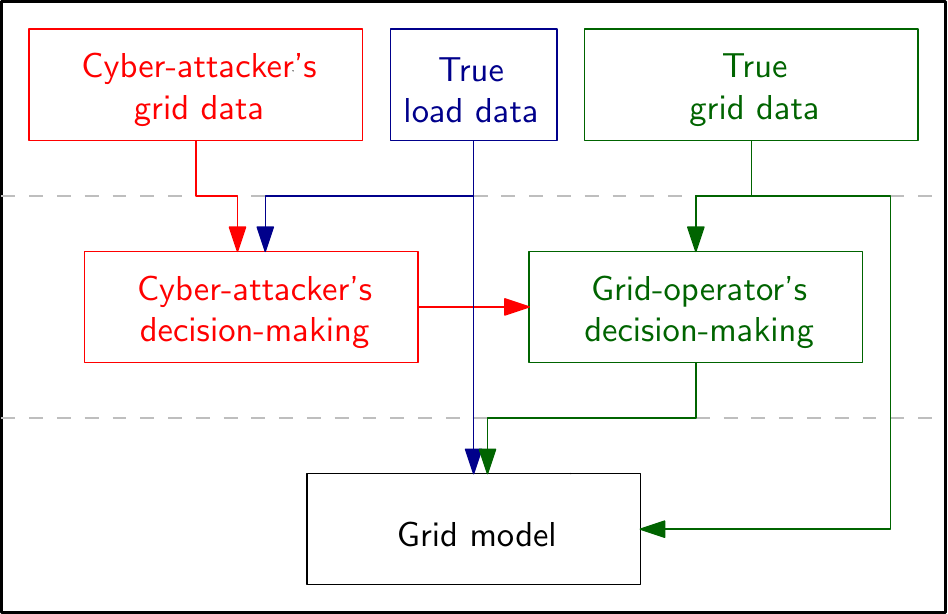}
\caption{Cyber-attack with imperfect information modeling}
\label{fig:overview}
\end{figure}

\subsection{Cyber-attack sequence modeling}
\label{attack_model}
Fig. \ref{fig:overview} presents the proposed flowchart for modeling a cyber-attack with imperfect information.

At the top layer of this figure we distinguish between two different datasets for the  grid, the so-called \emph{``cyber-attacker's''} and \emph{``true''} grid data. The former includes the (possibly inaccurate) data on the electricity grid that a cyber-attacker would exploit  to decide her cyber-attack strategy. The latter includes the correct values for all the parameters of the electricity grid and is only available to the grid-operator. Notice that the load demand data is not included in any of these two datasets as it is assumed to be a ``true'' set of data only known to the cyber-attacker.

The middle layer of Fig. \ref{fig:overview} models the interaction between the cyber-attacker and the grid-operator by means of two distinctive decision-making models, which are solved in sequence. First, the red box \emph{``cyber-attacker's decision-making''} corresponds to  bilevel model (\ref{up.obj} -- \ref{prat}) discussed in section \ref{blv}, used by the cyber-attacker to identify her attack vector for load redistribution (horizontal red arrow). The green box \emph{``grid-operator's decision-making''} models the reaction of the grid-operator to the cyber-attack. We must stress here that, even though a model for the grid-operator's reaction is embedded in the cyber-attacker's problem (\ref{up.obj} -- \ref{prat}), the true reaction of the grid-operator to the attack vector will be based on the true grid data she has access to. Therefore, to model such reaction, we solve here the lower-level optimization problem (\ref{low.obj} -- \ref{prat}) only,
given the optimal attack vector from the solution of (\ref{up.obj} -- \ref{prat}) and all parameters from the \emph{``true''} grid dataset\footnote{We should also acknowledge that restricting the physical models and equations in the grid-operator's decision-making model to match those of the cyber-attacker is not necessary by default. We made such choice here so as to isolate the impact of inaccurate data, and refer the reader to \cite{chu21} for a study of the impact of simplifications in the cyber-attacker's modeling of a grid-operator's decision-making.}.

The lower layer of Fig. \ref{fig:overview} illustrates a model of the physical impact of the cyber-attack on the electricity transmission grid, which is solved by combining: i) the true load data, ii) the true grid data and iii) the redispatching decisions the grid-operator would take given the load redistribution attack and her knowledge of the true grid data. Seeking to isolate the effect of imperfect information, in our implementation we combine such inputs through the same physical model as in the cyber-attacker's decision-making problem (\textit{i.e.}, the DC power flow equations) to measure grid insecurity in terms of the number and magnitude of overloaded branches\footnote{An alternative physical model, more detailed than the one used by the cyber-attacker, may well be relevant for generally assessing the system vulnerabilities as shown in \cite{liang2015}.}.

\subsection{Metrics related to cyber-physical risk assessment}
\label{assess_metrics}

Cyber-physical risk assessment serves to quantify the threat posed by a malicious cyber-attacker. The paradigm of the perfect information cyber-attack (\textit{i.e.}, a cyber-atacker having access to the \emph{``true''} grid data) is commonly employed in assessment applications, to anticipate the \emph{worst-case} physical impact on the electricity system. Acknowledging a cyber-attacker's imperfect information yields a set of random cyber-attack samples and respective impact indicators. Beyond the expected value and  distribution of the impact indicators over the Monte Carlo samples, we propose to analyze the risk of a cyber-attack with imperfect information by means of the following exclusive categories.

\begin{itemize}
\item{\emph{Perfect}: all samples wherein the attack vector of an imperfect cyber-attack matches the vector from the perfect information cyber-attack.}
\item{\emph{Success}: all other samples wherein an imperfect cyber-attack would still achieve the cyber-attacker's goals in terms of minimum number of overloaded branches with a flow above the respective threshold.}
\item{\emph{Partial success}: all other samples wherein an imperfect cyber-attack results in overloading at least one transmission branch with a flow above the respective threshold.}
\item{\emph{Failure}: all samples wherein an imperfect cyber-attack would cause no branch overload.}
\item{\emph{No attempt}: all samples wherein the cyber-attacker, given her imperfect information, fails to identify a feasible cyber-attack on the grid.}
\end{itemize}

The share of samples in the first category shows the relevance of the \emph{worst-case} perfect information cyber-attack, or alternatively the relevance of acknowledging a cyber-attacker's informational imperfections.  Note that this category does not
only include instances wherein the cyber-attacker's grid data randomly turn out to be perfectly accurate, but also instances wherein the cyber-attacker's informational imperfections have no effect on her strategy. A larger share of samples in the last two categories  {indicates} that the cyber-physical electricity system is inherently  {more} secure, either by way of ``absorbing'' the physical impact of imperfect cyber-attacks or by way of appearing more robust to the cyber-attacker. 

\subsection{Metrics related to cyber-physical risk management}
\label{mgmt_metrics}
Cyber-physical risk management serves to efficiently protect the grid from the threat of a malicious cyber-attacker. The perfect information cyber-attacker is commonly used in respective applications to anticipate a \emph{worst-case} atttack vector against which resources should be deployed in advance and/or prepared to be deployed. Facing a distribution of imperfect attackers translates into a distribution of attack vectors which we will classify by way of i) the assets in the cyber sub-system potentially targeted to launch  {(imperfect)} cyber-attacks, and ii) the assets of the physical sub-system that would undergo the physical impact of cyber-attacks. 

The first classification is more relevant for the deployment of preventive countermeasures on the cyber sub-system, in order to impede a successful cyber-physical attack. In the considered load redistribution attack mode, we propose to rank the system loads in terms of the share of attack vectors wherein their respective measurements are tampered with. In other words, rank the system loads in order of attack likelihood so as to efficiently select which load measurements to protect from falsifying. Noting that protecting (a sub-set of) the measurements under attack may suffice to render a load redistribution attack detectable, we will further count the share of imperfect attack vector samples that target an increasing sub-set (\textit{i.e.}, from at least one to all) of measurements in common with the perfect information cyber-attack.

We finally propose to rank groups of transmission grid branches in terms of the share of instances wherein all branches in a group would undergo an overload following a cyber-attack. Such ranking can be used to design effective emergency control strategies for the physical sub-system, so as to alleviate overloads in a timely manner before triggering cascading failure events.

\section{Case Studies}
\label{casestudy}

\subsection{Test case setup}
We adopt the single-area version (24 bus) of the IEEE-RTS96 benchmark\footnote{All system data can be found at \url{https://matpower.org/docs/ref/matpower5.0/case24_ieee_rts.html}.}. Following the practice of relevant studies (\textit{e.g.}, \cite{yuan2011,liang2015,tian2019}) we simulate a stressed operational condition by reducing  all branch transmission capacities to 65\% of the original values. We further model a malicious cyber-attacker seeking to overload at least $U=2$ transmission elements to at least $\rho_{\ell}=5\%$ of the respective capacities. We set the cyber-attacker's resource constraint to falsifying at most $A=10$ distinct load measurements and the maximum relative amount of false data per measurement to $\epsilon=20\%$. 

\subsection{Perfect information load redistribution attack}

\begin{figure}[h]
\centering
\includegraphics[width=0.5\textwidth]{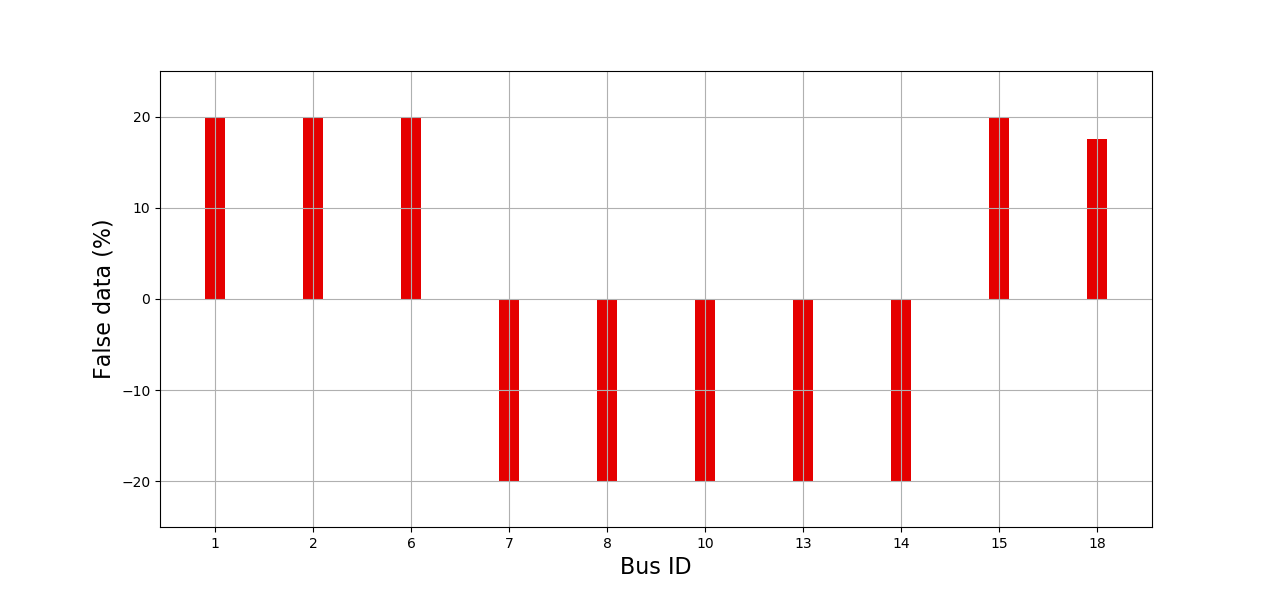}
\caption{Perfect information optimal attack vector}
\label{fig:perfect}
\end{figure}

Under the assumed conditions a cyber-attacker with perfect information,  solving model  (\ref{up.obj} -- \ref{prat}) with
the correct values for all grid parameters, would indeed be able to induce 2 overloads in the grid by more than 5\% of the respective branch capacities. More specifically, the cyber-attacker would provoke erroneous redispatch by the grid-operator eventually overloading branch 12 to 109.1\% of its capacity and branch 23 to 118.6\% of its capacity. The total magnitude of measurable overloads (\textit{i.e.}, above the 5\% threshold) would amount to 48.8 MW. Figure \ref{fig:perfect} illustrates the optimal attack vector, with the x-axis showing the index of the affected bus load meter and the y-axis the percentage of change in the falsified load data.

\subsection{Cyber-attacks with imperfect information on the grid admittances  {only}}

\begin{figure}[h]
\centering
\includegraphics[width=0.5\textwidth]{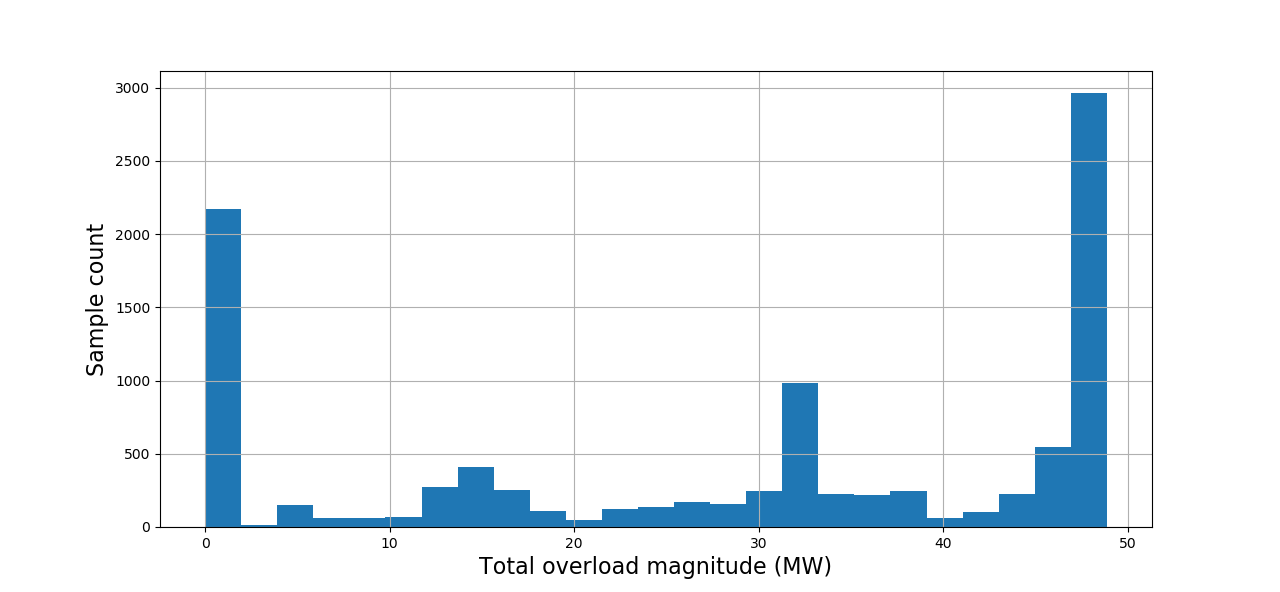}
\caption{Impact distribution of cyber-attacks with imperfect admittance data}
\label{fig:hist}
\end{figure}

We start by considering that a cyber-attacker may rely on inaccurate data considering the grid admittances only. To do so, we derive 10000 inaccurate grid samples, by applying a distinct error term to the admittance value of each branch, which is uniformly distributed in the range $\pm10\%$. Performing the respective simulations, we found that such (moderate) inaccuracy translates into 2677 (out of 10000) unique load redistribution attack vectors, with an average impact  {(\textit{i.e.}, total measurable overload)} of 28.36 MW. The histogram in Fig. \ref{fig:hist} shows the distribution of the impact of such potential attacks, which as anticipated ranges from 0 (for the case of not attempted or failed attacks) to the upper-bound set by the perfect information cyber-attack.

\begin{figure}[h]
\centering
\includegraphics[width=0.45\textwidth]{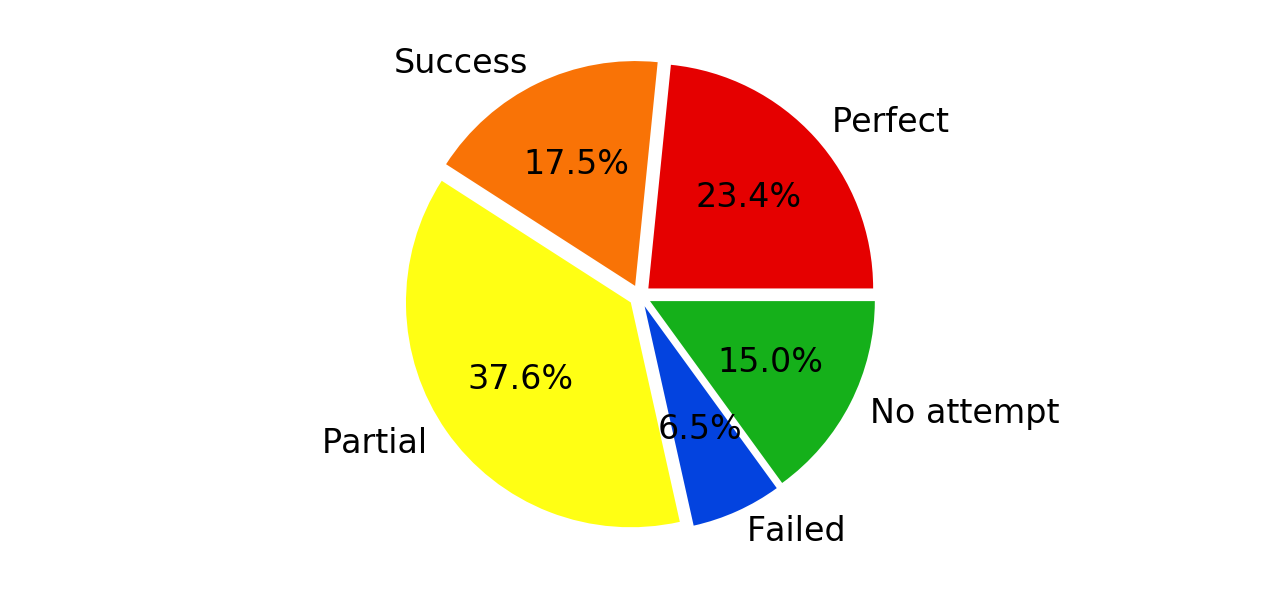}
\caption{Classification of cyber-attacks with imperfect admittance data}
\label{fig:classes}
\end{figure}

We further assess the risk posed by the imperfect cyber-attacker by means of the categories introduced in section \ref{assess_metrics} through the pie-chart in Fig. \ref{fig:classes}. As shown in this chart, due to the assumed informational imperfections the cyber-attacker would only be able to correctly identify the optimal perfect information attack vector from Fig. \ref{fig:perfect} on 23.4\% of the simulated instances. Conversely, on 15\% of the sampled instances the cyber-attacker would falsely believe that it would be fruitless to launch any load redistribution attack while on 6.5\% of the instances, she would launch an attack that would not be harmful to the grid.  Observing that on 40.9\% of the instances a cyber-attack with imperfect information would cause an overflow on at least two grid branches, while on 78.5\% it would cause an overflow on at least one branch, we may infer from Fig. \ref{fig:classes}
that this system is insecure\footnote{One may notice however that informational imperfections are in favor of security, as a perfectly informed attacker would be able to induce insecurity with 100\% likelihood.}. 

\begin{figure}[h]
\centering
\includegraphics[width=0.5\textwidth]{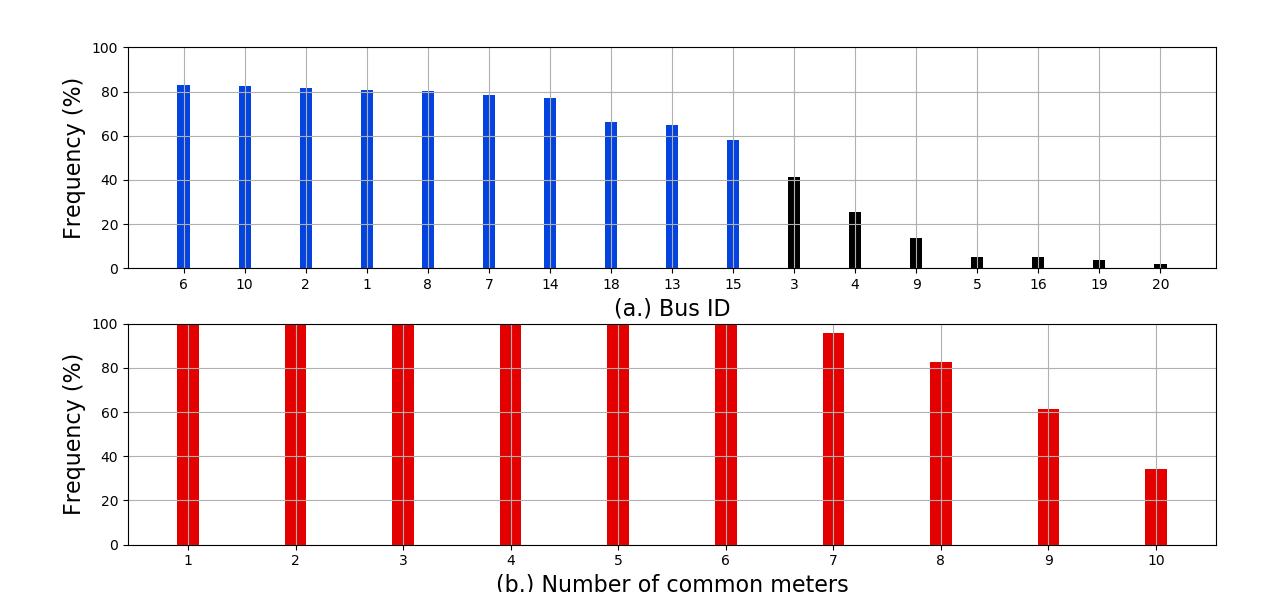}
\caption{Frequency of attacks (a.)  per meter and (b.) sharing common meters with the perfect information attack }
\label{fig:meters}
\end{figure}

Pursuing the analysis from a risk management perspective, Fig. \ref{fig:meters}.a shows the relative frequency of attacking each distinct bus load meter amongst the 10000 sampled cyber-attacks. The bars in blue correspond to the perfect information optimal attack vector from Fig. \ref{fig:perfect} and it is notable that these are the meters ranked first in order of decreasing frequency. Indeed, as illustrated further in Fig. \ref{fig:meters}.b, 100\% of the imperfect cyber-attacks share at least one (in fact, even at least five) common attacked asset(s) with the perfect information cyber attack while all 10 meters from Fig. \ref{fig:perfect} have been attacked in  39.5\% of the sampled instances. The important take-away here is that protecting the meters  {that would have been attacked in the perfect information case} may well be sufficient to detect and prevent  {with very high probability the cyber-attacks under imperfect information} from physically harming the system.

\begin{figure}[h]
\centering
\includegraphics[width=0.5\textwidth]{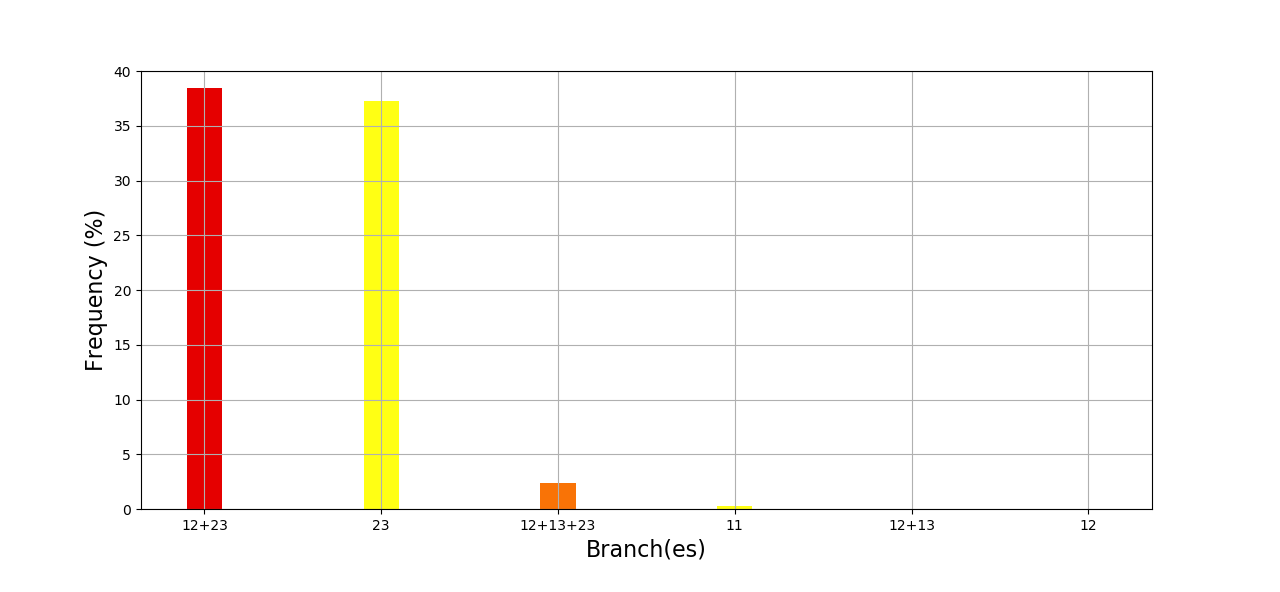}
\caption{Physical impact of of cyber-attacks with imperfect admittance data}
\label{fig:physical}
\end{figure}

Finally, Fig.\ref{fig:physical} demonstrates which transmission branches would be overloaded due to the imperfect cyber-attacks. Adopting the color-coding of Fig. \ref{fig:classes}, we show that for a large share of the samples the  imperfect cyber-attack results in overloading the same branches as the perfect information attack, albeit to a smaller degree. The take-away here is that taking physical preventive/corrective measures for the possible joint outage of these branches could also be an effective strategy for managing cyber-physical risk. Notice the small frequency of imperfect cyber-attacks overloading three branches, which are suboptimal in terms of  total overload magnitude.

\subsection{Sensitivity analysis with respect to the admittance error range}
\begin{figure}[h]
\centering
\includegraphics[width=0.5\textwidth]{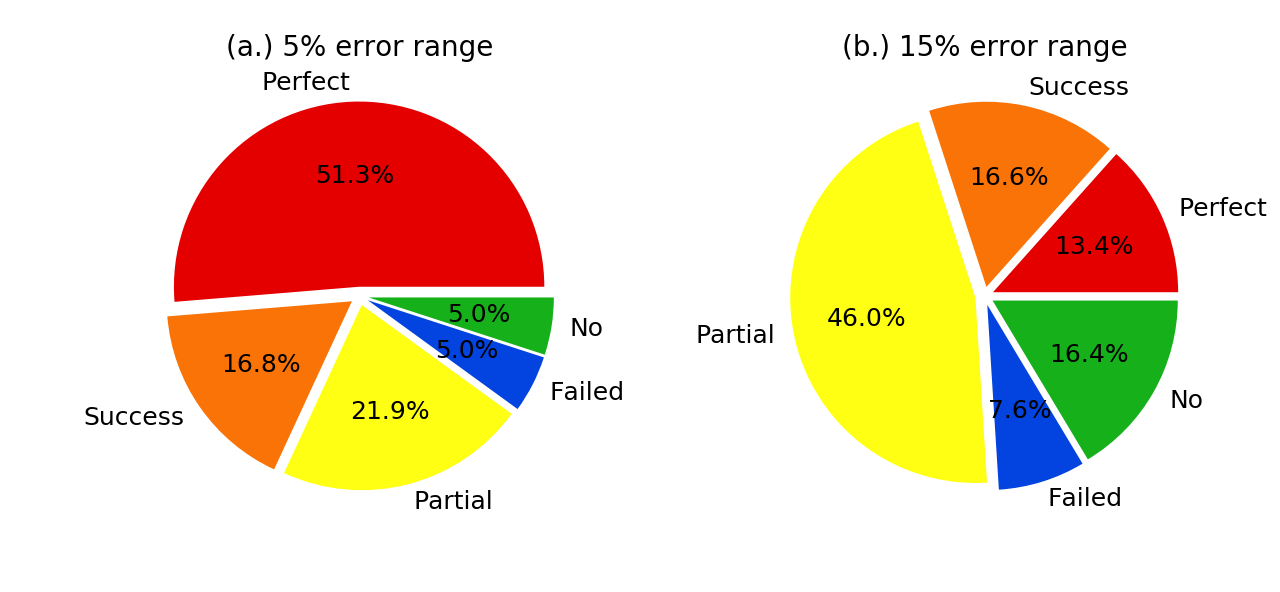}
\caption{Classifications for (a.) $\pm5\%$ and (b.) $\pm15\%$ admittance error}
\label{fig:two_pies}
\end{figure}

To validate the aforementioned observations we perform a sensitivity analysis by drawing two additional samples of 10000 inaccurate grid instances while assuming that the imperfect cyber-attacker's error in admittance values is uniformly distributed in the $\pm5\%$ and $\pm15\%$ ranges. As anticipated, in the former case the average impact of the imperfect cyber-attacks increases to 35.6 MW  (with 1428 unique attack vectors) while in the latter it reduces slightly to 26.72 MW (with 4044 unique attack vectors).  It is noteworthy that in the case of reduced inaccuracy, Fig. \ref{fig:two_pies}.a., the percentage of so-called \emph{perfect} attacks more-than doubles to 51.3\%. This shows that (the reduced) inaccuracy has a smaller effect on the attack vector of the imperfect cyber-attacker. Conversely, in
Fig. \ref{fig:two_pies}.b., increased inaccuracy almost halves the percentage of \emph{perfect attacks}, with the most notable increase observed in the \emph{partial} attack class.

In our detailed results we further find that for both cases (\textit{i.e.}, under reduced or increased randomness) the set of meters included in the optimal perfect information attack vector from Fig. \ref{fig:perfect} remains the set of the 10 most frequently attacked meters, while the frequency of attacking at least one of these meters remains as high as 100\%. These findings are well in line with the argument that protecting the meters involved in the perfect information cyber-attack is a good starting point for detecting and preventing any random  imperfect cyber-attack vector. Similarly, concerning the branches that may undergo overloads in the aftermath of an imperfect cyber-attack, our sensitivity analysis detailed results qualitatively follow the representation of  {Fig.} \ref{fig:physical}. That is, most frequently both branches that would be overloaded in the case of the perfect cyber-attack are also affected by the imperfect cyber-attacks.

\subsection{Cyber-attacks with imperfect information on the  {branch capacities only}}

We continue the analysis by henceforth considering the case  {where the cyber-attacker relies} on inaccurate data  {about} the branch capacities only. We sample additionally 10000 inaccurate grids, by applying a distinct error term to the capacity value of each branch, which is again uniformly distributed in the range $\pm10\%$.  With such assumptions, the average cyber-attack impact reduces to 25.31 MW while the number of unique cyber-attacks increases to 6737.
\begin{figure}[h]
\centering
\includegraphics[width=0.45\textwidth]{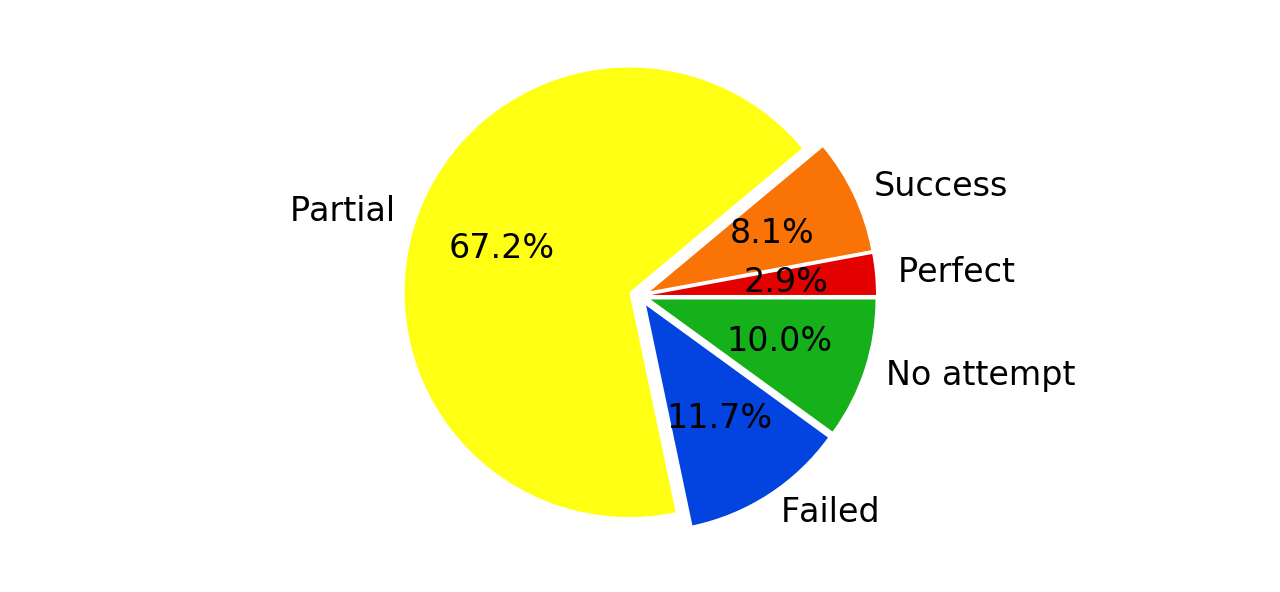}
\caption{Classification of cyber-attacks with imperfect capacity data}
\label{fig:cap_classes}
\end{figure}

Fig. \ref{fig:cap_classes} presents the classification of the random cyber-attacks as per the categories introduced in section \ref{assess_metrics}. The qualitative difference with respect to imperfect admittance values is striking in comparison to Fig. \ref{fig:classes}. Indeed, for the same error range: i) the share of \emph{perfect} attacks has collapsed, ii) the share of \emph{success} attacks is  {(more than)} halved, iii) the share of \emph{partial} attacks is considerably increased, and, iv) the share of ineffective attacks is moderately increased.  In other words, imperfect information on the branch capacities leads to much less effective cyber-attacks posing a smaller risk to the system cyber-physical security.

We can identify systematic reasons for this finding. Indeed, in case the cyber-attacker undervalues branch capacities, she is prone to overestimating the impact of an attack vector in firstly misleading the grid-operator to redispatch generation to avoid overloads under the load redistribution, and secondly in causing actual overloads by way of the erroneous redispatch. This explains the large shift from \emph{perfect/success} to \emph{partial} attacks.  Also, in case the cyber-attacker overvalues branch capacities, she is prone to believing there is no potential for attacking the grid.

\begin{figure}[h]
\centering
\includegraphics[width=0.5\textwidth]{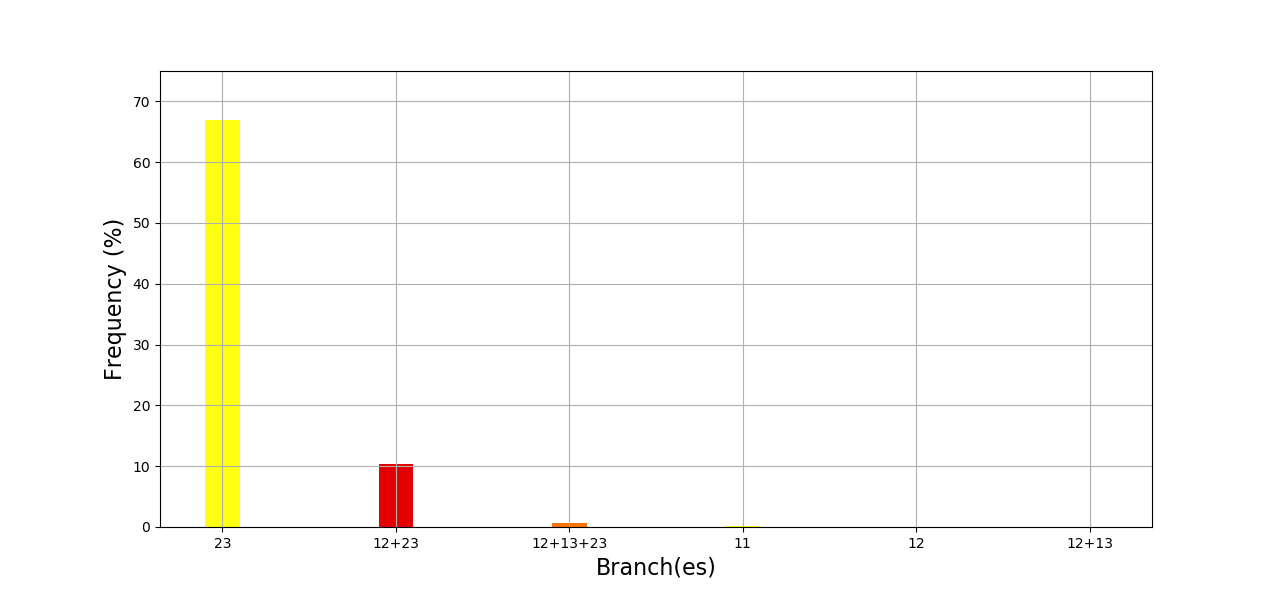}
\caption{Physical impact of of cyber-attacks with imperfect capacity data}
\label{fig:physical_rates}
\end{figure}

Concerning risk management, we once again find that the frequency of attacking at least one meter in common with the perfect information attack remains 100\%. As should be anticipated by the dominance of the \emph{partial} attack category, the most frequent overflow in the system now concerns a single transmission branch, Fig. \ref{fig:physical_rates}. Notice here that the groups of affected branches (x-axis) are all in common with Fig. \ref{fig:physical}.  Both these findings further showcase the relevance of these groups of cyber and physical  sub-system assets for preventive and corrective cyber-physical risk management. 

\subsection{Computational environment}

Our implementation of the Monte Carlo simulation framework was developed in Julia \cite{julia} using the JuMP modeling language \cite{jump} and the PowerModels.jl framework \cite{PowerModels}. We solved  {the} bilevel optimization problem  (\ref{up.obj} -- \ref{prat}) \textit{via} its single level  equivalent reformulation, replacing the lower-level inner minimization problem (\ref{low.obj} -- \ref{prat}) with its \emph{Karush-Kuhn-Tucker} optimality conditions.   We used the \emph{big-M} approach to rewrite the disjunctive inequalities expressing the complementary slackness conditions as mixed-integer linear constraints and eventually solved all instances of the single-level MILP problem with the CPLEX \cite{cplex} solver.

\section{Conclusions}
\label{theend}

In this paper we  {have} modeled the  {risk of cyber-physical insecurities} of the electricity transmission grid,  {while explicitly taking into account plausible informational imperfections of a real-world cyber-attacker}. We   {have} considered the scenario of a load redistribution attack with the final aim of inducing measurable overloads (\textit{i.e.}, beyond a minimum relative magnitude) to a number of transmission branches. For the purposes of this study, we have introduced novel constraint expressions to reflect a cyber-attacker's intention to create an overwhelming grid insecurity in the standard $\maxmin$ load redistribution problem formulation. We further   {have} performed a series of Monte Carlo simulations, modeling potential cyber-attackers with inaccurate data on branch addmittances or transmission capacities, and   {we have} proposed a set of metrics to synthesize the outcome of such simulations in the context of risk assessment and risk management. 

From a risk  {\emph{assessment}} perspective, we have found  {in our case study} that inaccurate knowledge of the grid admittance matrix is not a considerable impediment to inducing physical insecurity through the cyber sub-system. Indeed, for an increasing degree of inaccuracy on the branch admittance values between $[5 - 15]\%$ the frequency of cyber-attacks putting the IEEE-RTS96 benchmark in an insecure state was found in the $[90 - 76]\%$ range.  {On the other hand,} relying on imperfect information on the branch transmission capacities  {was found}   {to lead to a quite stronger reduction of the} cyber-physical risk as it may lead a cyber-attacker to either i) launch less effective attacks  {when underestimating some} branch capacities, and/or ii)  {give up the idea of} attacking the system  {when overestimating some of them}.

From a risk   {\emph{management}} perspective, we most notably  {observed in our case study} that in spite of random inaccuracies, 100\% of the random attack vectors would target at least one meter in common with the  {``ideal''} perfect information attack. This implies that monitoring the state of the meters that a perfectly informed attacker (\textit{i.e.}, the  {``worst-case'' from the view-point of the electricity grid end-users}) would select could be a very effective preventive detection strategy. Moreover the set of the grid assets undergoing the physical impact of the cyber-attack was in all cases found to be relatively small, opening the possibility for efficient attack mitigation strategies on the physical sub-system.

Notice that while we relied on the specific load redistribution scenario and the specific cyber-attacker model introduced here, our analysis in principle generalizes to alternative cyber-attack instances, provided that the cyber-attacker is indeed optimizing her strategy while presuming perfect knowledge of the grid model and the grid-operator's strategy. Further work will therefore be devoted in modeling alternative cyber-attacker types, for instance an actor potentially launching any attack vector that meets some impact threshold constraints (\textit{i.e.}, any feasible rather than an optimal solution to a bilevel optimization model). Beyond this direction, we will also pursue the question of efficiently taking cyber-physical risk management decisions under uncertainty on the realistic cyber-attacker properties.

\bibliographystyle{IEEEtran}
\bibliography{full_eklw}

\begin{thebibliography}{10}
\providecommand{\url}[1]{#1}
\csname url@samestyle\endcsname
\providecommand{\newblock}{\relax}
\providecommand{\bibinfo}[2]{#2}
\providecommand{\BIBentrySTDinterwordspacing}{\spaceskip=0pt\relax}
\providecommand{\BIBentryALTinterwordstretchfactor}{4}
\providecommand{\BIBentryALTinterwordspacing}{\spaceskip=\fontdimen2\font plus
\BIBentryALTinterwordstretchfactor\fontdimen3\font minus
  \fontdimen4\font\relax}
\providecommand{\BIBforeignlanguage}[2]{{%
\expandafter\ifx\csname l@#1\endcsname\relax
\typeout{** WARNING: IEEEtran.bst: No hyphenation pattern has been}%
\typeout{** loaded for the language `#1'. Using the pattern for}%
\typeout{** the default language instead.}%
\else
\language=\csname l@#1\endcsname
\fi
#2}}
\providecommand{\BIBdecl}{\relax}
\BIBdecl

\bibitem{kirschen2008}
D.~Kirschen and F.~Bouffard, ``Keeping the lights on and the information
  flowing,'' \emph{IEEE Power and Energy Magazine}, vol.~7, no.~1, pp. 50--60,
  2009.

\bibitem{Zhang2021}
H.~Zhang, B.~Liu, and H.~Wu, ``Smart grid cyber-physical attack and defense: A
  review,'' \emph{IEEE Access}, vol.~9, pp. 29\,641--29\,659, 2021.

\bibitem{he2016}
\BIBentryALTinterwordspacing
H.~He and J.~Yan, ``Cyber-physical attacks and defences in the smart grid: a
  survey,'' \emph{IET Cyber-Physical Systems: Theory \& Applications}, vol.~1,
  pp. 13--27(14), December 2016. [Online]. Available:
  \url{https://digital-library.theiet.org/content/journals/10.1049/iet-cps.2016.0019}
\BIBentrySTDinterwordspacing

\bibitem{liu2011}
\BIBentryALTinterwordspacing
Y.~Liu, P.~Ning, and M.~K. Reiter, ``False data injection attacks against state
  estimation in electric power grids,'' vol.~14, no.~1, Jun. 2011. [Online].
  Available: \url{https://doi.org/10.1145/1952982.1952995}
\BIBentrySTDinterwordspacing

\bibitem{yuan2011}
Y.~Yuan, Z.~Li, and K.~Ren, ``Modeling load redistribution attacks in power
  systems,'' \emph{IEEE Transactions on Smart Grid}, vol.~2, no.~2, pp.
  382--390, 2011.

\bibitem{liang2015}
J.~Liang, L.~Sankar, and O.~Kosut, ``Vulnerability analysis and consequences of
  false data injection attack on power system state estimation,'' \emph{IEEE
  Transactions on Power Systems}, vol.~31, no.~5, pp. 3864--3872, 2015.

\bibitem{zhang2016}
J.~Zhang and L.~Sankar, ``Physical system consequences of unobservable
  state-and-topology cyber-physical attacks,'' \emph{IEEE Transactions on Smart
  Grid}, vol.~7, no.~4, 2016.

\bibitem{tian2019}
M.~Tian, M.~Cui, Z.~Dong, X.~Wang, S.~Yin, and L.~Zhao, ``Multilevel
  programming-based coordinated cyber physical attacks and countermeasures in
  smart grid,'' \emph{IEEE Access}, vol.~7, pp. 9836--9847, 2019.

\bibitem{rahman2012}
M.~A. Rahman and H.~Mohsenian-Rad, ``False data injection attacks with
  incomplete information against smart power grids,'' in \emph{2012 IEEE Global
  Communications Conference (GLOBECOM)}, 2012, pp. 3153--3158.

\bibitem{zhang2018}
J.~Zhang, Z.~Chu, L.~Sankar, and O.~Kosut, ``Can attackers with limited
  information exploit historical data to mount successful false data injection
  attacks on power systems?'' \emph{IEEE Transactions on Power Systems},
  vol.~33, no.~5, pp. 4775--4786, 2018.

\bibitem{anibal}
A.~Sanjab and W.~Saad, ``On bounded rationality in cyber-physical systems
  security: Game-theoretic analysis with application to smart grid
  protection,'' in \emph{2016 Joint Workshop on Cyber- Physical Security and
  Resilience in Smart Grids (CPSR-SG)}, 2016, pp. 1--6.

\bibitem{test_system}
C.~Grigg, P.~Wong, P.~Albrecht, R.~Allan, M.~Bhavaraju, R.~Billinton, Q.~Chen,
  C.~Fong, S.~Haddad, S.~Kuruganty, W.~Li, R.~Mukerji, D.~Patton, N.~Rau,
  D.~Reppen, A.~Schneider, M.~Shahidehpour, and C.~Singh, ``{The IEEE
  Reliability Test System-1996. A report prepared by the Reliability Test
  System Task Force of the Application of Probability Methods Subcommittee},''
  \emph{IEEE Transactions on Power Systems}, vol.~14, no.~3, pp. 1010--1020,
  Aug 1999.

\bibitem{che2018false}
L.~Che, X.~Liu, Z.~Li, and Y.~Wen, ``False data injection attacks induced
  sequential outages in power systems,'' \emph{IEEE Transactions on Power
  Systems}, vol.~34, no.~2, pp. 1513--1523, 2018.

\bibitem{chu21}
Z.~Chu, J.~Zhang, O.~Kosut, and L.~Sankar, ``N-1 reliability makes it difficult
  for false data injection attacks to cause physical consequences,'' \emph{IEEE
  Transactions on Power Systems}, vol.~36, no.~5, pp. 3897--3906, 2021.

\bibitem{julia}
\BIBentryALTinterwordspacing
J.~Bezanson, A.~Edelman, S.~Karpinski, and V.~Shah, ``Julia: A fresh approach
  to numerical computing,'' \emph{SIAM Review}, vol.~59, no.~1, pp. 65--98,
  2017. [Online]. Available: \url{https://doi.org/10.1137/141000671}
\BIBentrySTDinterwordspacing

\bibitem{jump}
\BIBentryALTinterwordspacing
I.~Dunning, J.~Huchette, and M.~Lubin, ``{JuMP}: A modeling language for
  mathematical optimization,'' \emph{{SIAM} Review}, vol.~59, no.~2, pp.
  295--320, 2017. [Online]. Available: \url{https://doi.org/10.1137/15M1020575}
\BIBentrySTDinterwordspacing

\bibitem{PowerModels}
C.~Coffrin, R.~Bent, K.~Sundar, Y.~Ng, and M.~Lubin, ``{PowerModels.jl}: An
  open-source framework for exploring power formulations,'' in \emph{2018 Power
  Systems Computation Conference {(PSCC)}}, June 2018.

\bibitem{cplex}
I.~I. Cplex, ``V12. 1: User’s manual for {CPLEX},'' \emph{International
  Business Machines Corporation}, vol.~46, no.~53, p. 157, 2009.

\end{thebibliography}

\end{document}